\documentclass[a4paper]{article}

\usepackage{icrc2013}

\title{Time-dependent cosmic ray modulation in the outer heliosphere: Signatures of a heliospheric asymmetry and model predictions along Voyager 1 and 2 trajectories}

\shorttitle{Cosmic ray modulation in the outer heliosphere}

\authors{
R. Manuel$^{1}$,
S.E.S. Ferreira$^{1}$,
M.S. Potgieter$^{1}$
}

\afiliations{
$^1$ Centre for Space Research, North-West University, Potchefstroom 2520, South Africa.
}

\email{rexmanuel@live.com}

\abstract{A two-dimensional, time-dependent numerical model is used to calculate the modulation of cosmic rays in the heliosphere. Computations are compared to spacecraft observations in the inner and outer heliosphere. It is shown that the model produces cosmic ray proton intensities compatible to different spacecraft observations on a global scale, at Earth and along both Voyager spacecraft trajectories. The study reveals that when the same modulation parameters, which resulted in compatible intensities along Voyager 1, were assumed along the Voyager 2 trajectory, the model failed to reproduce the observations. The study also found that any change in diffusion parameters alone could not reproduce the cosmic ray observations along Voyager 2 so that changes to the heliospheric geometry were necessary i.e the computed intensities along both Voyager trajectories suggest that the heliosphere is asymmetric. Furthermore, $E$ $>$ 70 MeV and 133-242 MeV proton intensities along Voyager 1 and 2 trajectories are predicted from end of 2012 onwards. It is shown that the computed intensities along Voyager 1 increase with an almost constant rate up to the heliopause. However, the model shows that Voyager 2 is still under the influence of temporal solar activity changes because of its relatively large distance to the heliopause. Along the Voyager 2 trajectory, the intensities remained generally constant for some time and should soon start to increase steadily.}


\begin{document}
\maketitle

\section{Introduction}
Galactic cosmic ray (CR) modulation along Voyager 1 (V1) and Voyager 2 (V2) trajectories are computed using a 2D time-dependent modulation model and compared to $E$ $>$ 70 MeV and 133-242 MeV proton observations.  Recent theoretical advances in transport coefficients by \cite{Shalchi-etal-2004}, \cite{Teufel-and-Schlickeiser-2002},  \cite{Teufel-and-Schlickeiser-2003} and \cite{Minnie07} are implemented in the model. The measured magnetic field magnitude, variance and tilt angle are transported from Earth into the heliosphere to provide a time-dependence for the transport parameters. It is shown that the model computed compatible CR intensities at Earth and along both the Voyager trajectories when compared to the spacecraft observations. 

The model results confirm that different transport parameters along the V1 and V2 trajectories are not sufficient to reproduce the CR observations. A heliospheric asymmetry in the assumed heliospheric geometry is necessary. Such an asymmetry was already proposed by MHD models by \cite{Opher-etal-2009} and \cite{Pogorelov-etal-2009} due to an external pressure resulting from the interstellar magnetic field (see also \cite{Ngobeni-2011}). 

CR intensities along both Voyager trajectories are predicted up to the heliopause (HP). The computed results show that the V1 intensities increase at a constant rate up to the HP, but V2 intensities should show the influence of temporal changes in solar activity due to the large distance to the HP compared to V1. 

\section{Model}
The 2D time-dependent numerical model (see \cite{Potgieter-and-Leroux-1992},\cite{Ferreira-and-Potgieter-2004}) is based on solving the Parker transport equation \cite{Parker65}: 
\begin{eqnarray}\label{tpe}
 \frac{\partial f}{\partial t}=&
-\left( \vec{V} + \left\langle \vec{v_{D}} \right\rangle \right)  \cdot\nabla f +\nabla\cdot(\mathbf{K_{S}} \cdot\nabla 
 f) \nonumber \\ 
 &+ \frac{1}{3}(\nabla\cdot \vec{V})\frac{\partial f}{\partial \ln P} + Q.
\end{eqnarray}

Here $t$ is the time, $\vec{V}$ is the solar wind velocity,  $\left\langle \vec{v_{D}} \right\rangle$ the pitch angle averaged guiding center drift velocity for a near isotropic distribution function $f$, $\mathbf{K_{S}}$ is the isotropic diffusion tensor, $P$ is rigidity and $Q$ is any particle source inside the heliosphere. This equation is solved numerically in terms of $t$ and $P$ in two-dimensional space $(r, \theta)$ with $r$ radial distance and $\theta$ polar angle. 

\begin{figure*}[t]
\begin{center}
\includegraphics [width=0.64\textwidth]{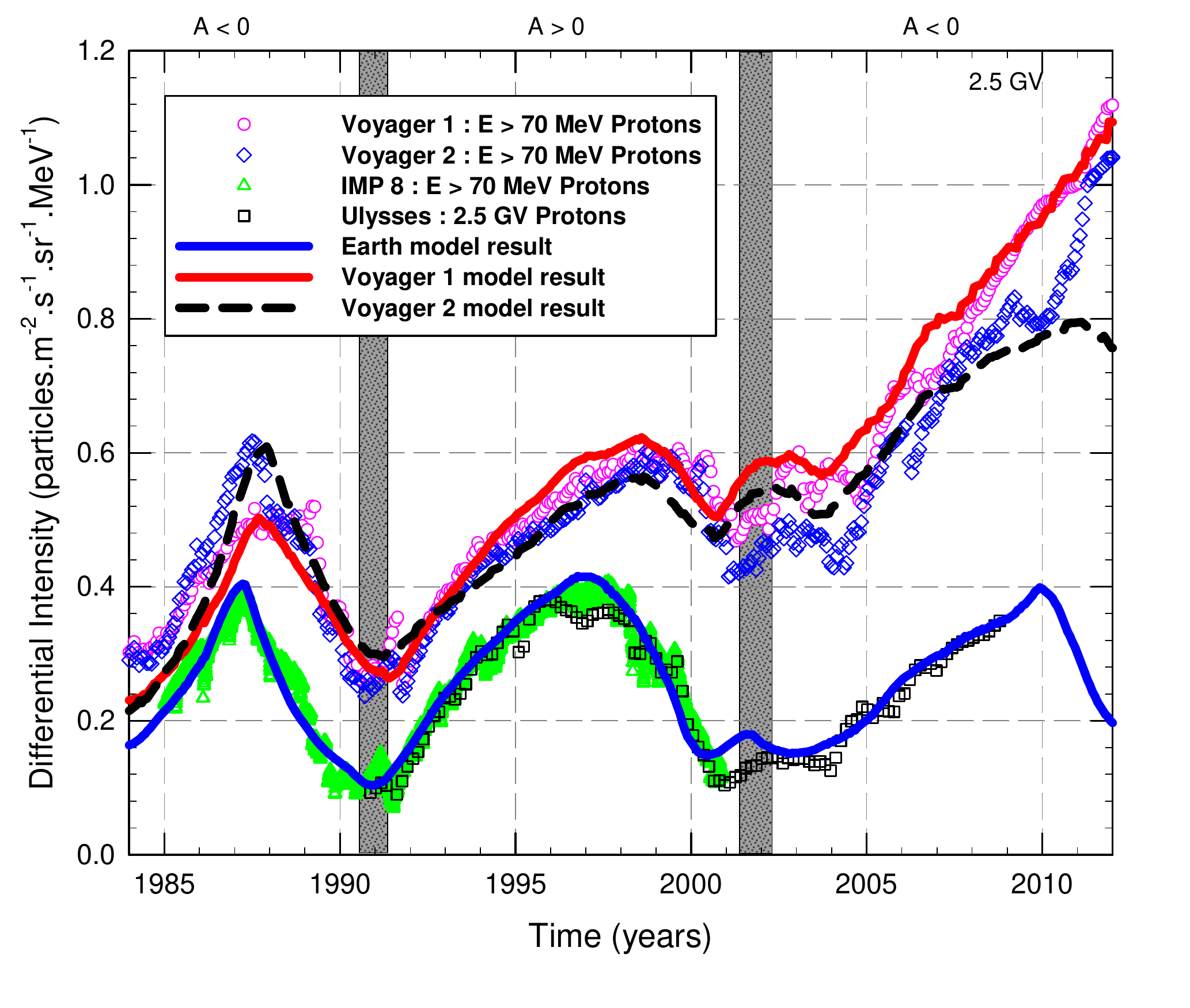}
\end{center}
\caption{Proton observations  (symbols) are shown as a function of time for V1, V2, IMP 8 and Ulysses. Also shown are the 2.5 GV model results at Earth and along the V1 and V2 trajectories. From \cite{Manuel13}.}
\label{figure1}
\end{figure*}

At the energies pertinent to this study we focus on two important transport processes, diffusion and drift. The corresponding diffusion coefficients in the radial direction ($K_{rr}$), the polar direction ($K_{\theta\theta}$) and the drift coefficient ($K_{A}$) are respectively,
\begin{eqnarray}\label{DiffCoefs}
K_{rr}&=&K_{||}\cos^2\psi+K_{\bot r}\sin^2\psi, \\
K_{\theta\theta}&=&K_{\bot\theta},\\
K_{A}&=&\frac{\beta P}{3B}\frac{10P^{2}}{10P^{2}+1},
\end{eqnarray}
where $K_{||}$ is the diffusion coefficient parallel to the HMF, $K_{\bot r}$ the perpendicular diffusion coefficient in the radial direction and $K_{\bot\theta}$ the perpendicular diffusion coefficient in the polar direction respectively.  Also $B$ is the HMF magnitude, $\psi$ is the spiral angle of $B$ and $\beta$ the ratio between the particle speed to the speed of light. For an illustration of the dependence of these coefficients on $r, \theta$ and $P$, see \cite{Manuel11}.  

This study assumes rigidity dependence for $K_{||}$  as calculated by \cite{Teufel-and-Schlickeiser-2002} for protons (damping model) in the inner heliosphere, 
\begin{eqnarray}\label{par}
\lambda_{||}=C_{1}\left( \frac{P}{P_{0}}\right) ^{1/3} \left( \frac{r}{r_{0}}\right)^{C_{2}} f_{2}(t)
\end{eqnarray}
where $C_{1}$ is a constant with units of AU, $P_{0} =$ 1 MV, $r_{0}=1$ AU,  $C_{2}$ a constant and $f_{2}(t)$ a time-dependent function.

For perpendicular diffusion coefficient we assume, 
\begin{eqnarray}\label{perp1}
K_{\bot r} = a K_{||} \frac{f_{3}(t)}{f_{2}(t)} \\ \label{perp2}
K_{\bot \theta} = b K_{||} F(\theta)\frac{f_{3}(t)}{f_{2}(t)}
\end{eqnarray}

with $a=0.022$, $b=0.01$, $F(\theta)$ a function enhancing $K_{\bot \theta}$ toward the poles by a factor of 6 and $f_{3}(t)$ a time-varying function.

The theoretical advances in transport parameters by \cite{Shalchi-etal-2004}, \cite{Teufel-and-Schlickeiser-2002},  \cite{Teufel-and-Schlickeiser-2003} and \cite{Minnie07} are incorporated into our time-dependent transport model to compute the time-dependence for the transport parameters. The magnetic field magnitude $B$, magnetic field variance $\delta{B^{2}}$  and tilt angle are transported from Earth into the outer heliosphere resulting in a time-dependence for the diffusion parameters. 

The time dependence for $K_{||}$, the diffusion coefficient parallel to the HMF, is attained from an expression for parallel free mean path $\lambda_{||}$ for protons given by \cite{Teufel-and-Schlickeiser-2003} and since we consider only the influence of time varying quantities $B$ and $\delta{B^{2}}$ on $\lambda_{||}$, we approximate the complicated equation (see also \cite{Manuel11b}) and the time dependence of $K_{||}$ is then given by,
\begin{eqnarray}\label{f2}
f_{2}(t)= C_{4}\left( \frac{1}{\delta{B}(t)}\right)^2 
\end{eqnarray}
where $C_{4}$ is a constant in units of $\textrm{(nT)}^2$.

And for $f_{3}(t)$, the time dependence of perpendicular diffusion coefficients, we approximate the expression for $\lambda_{\bot}$ as given by \cite{Shalchi-etal-2004} as:
\begin{eqnarray}\label{f3} 
f_{3}(t)=C_{5}\left(\frac{\delta{B}(t)}{B(t)}\right)^\frac{4}{3}\left(\frac{1}{\delta B(t)}\right)^\frac{2}{3}
\end{eqnarray}
where $C_{5}$ a constant in units of $\textrm{(nT)}^{2/3}$.

 \begin{figure}[t]
  \centering
  \includegraphics[width=0.5\textwidth]{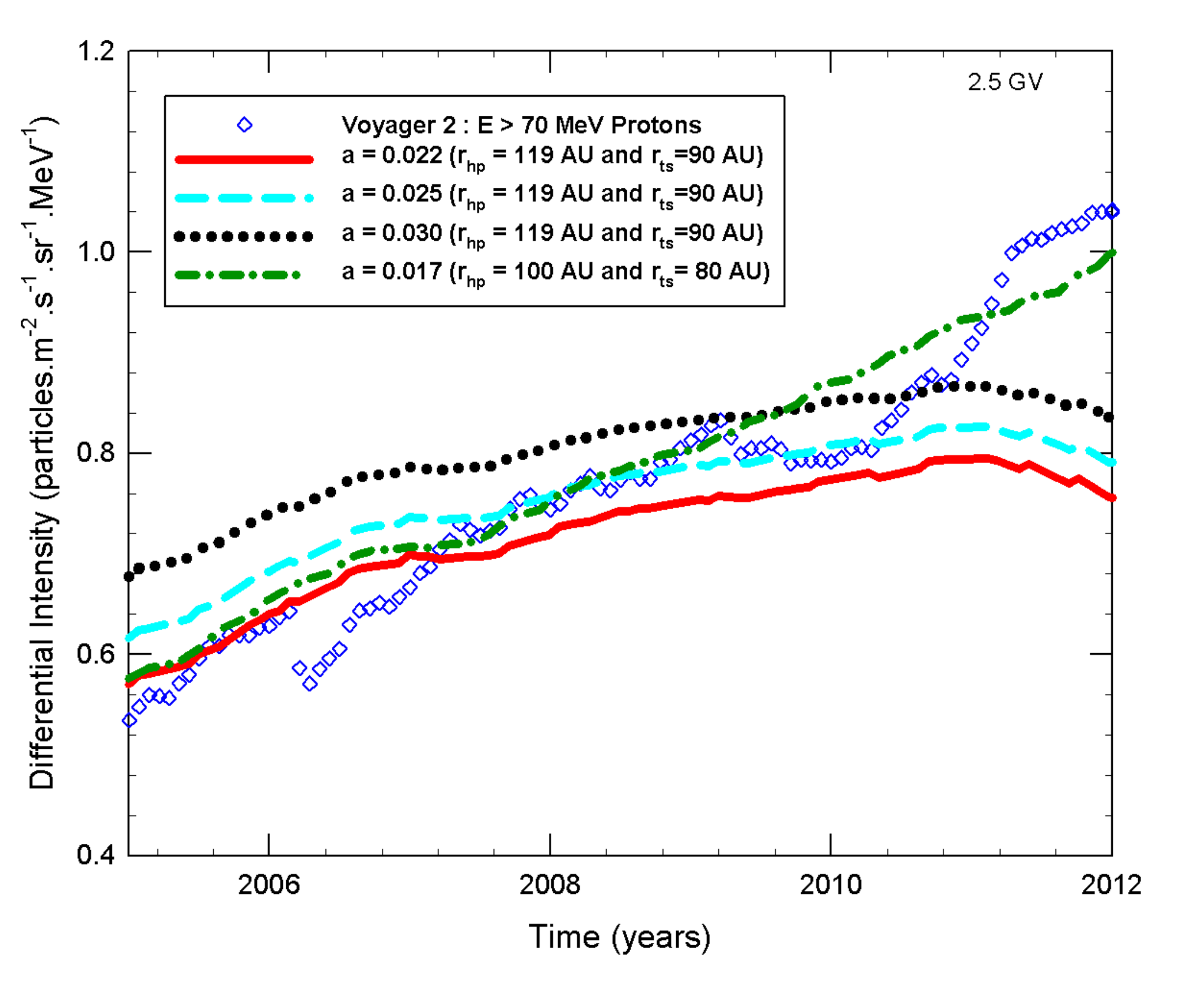}
  \caption{Similar to Figure \ref{figure1} except that here modelling results along the V2 trajectory are shown for different $a$ values, $r_{hp}$ and $r_{ts}$.}
  \label{figure2}
 \end{figure}
 
 \begin{figure}[t]
  \centering
  \includegraphics[width=0.48\textwidth]{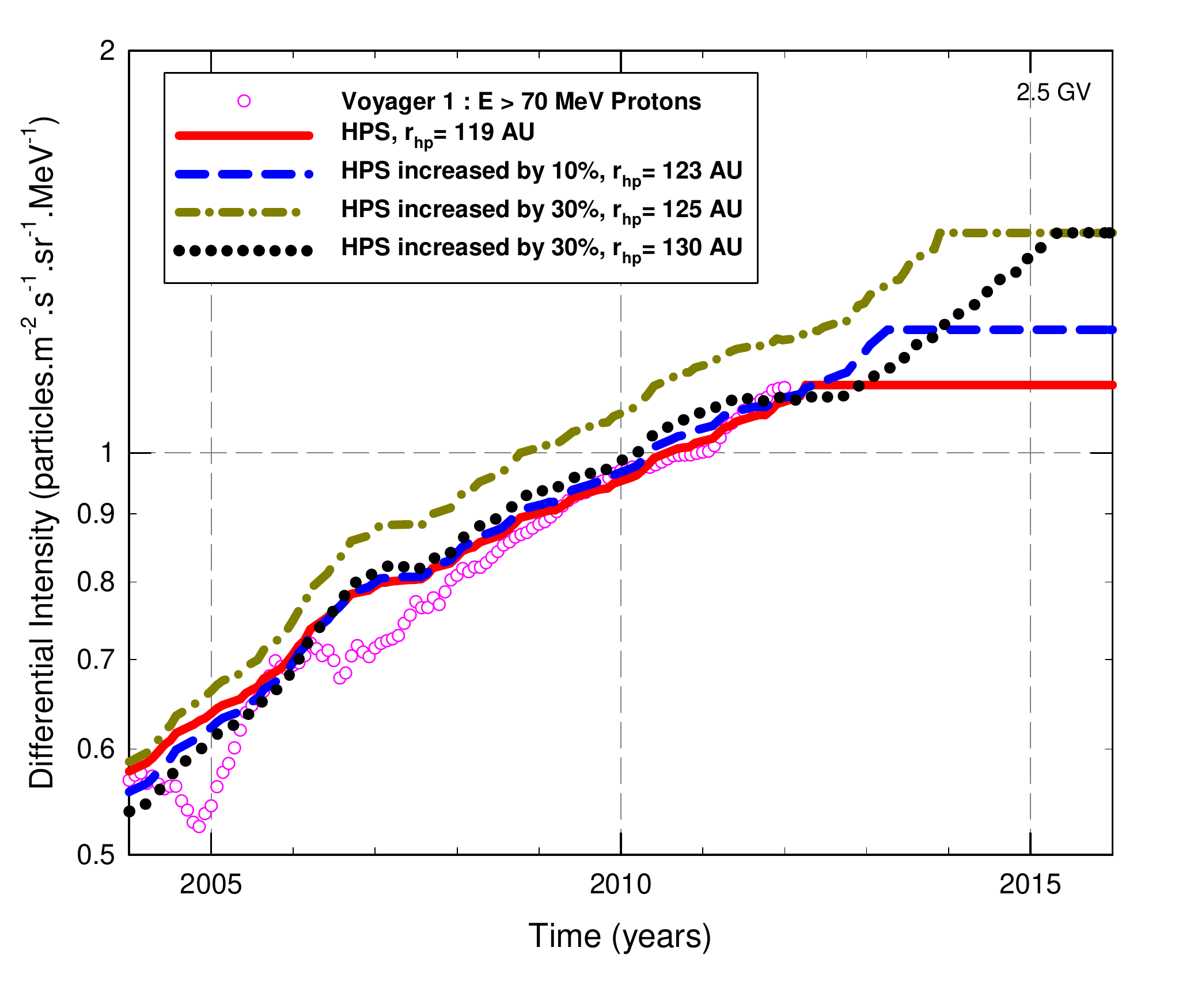}
  \caption{Similar to Figure \ref{figure1} except that here modelling results along the V1 trajectory are shown for different assumed HPS values and $r_{hp}$.}
  \label{figure3}
 \end{figure}

\begin{figure}[t]
  \centering
  \includegraphics[width=0.48\textwidth]{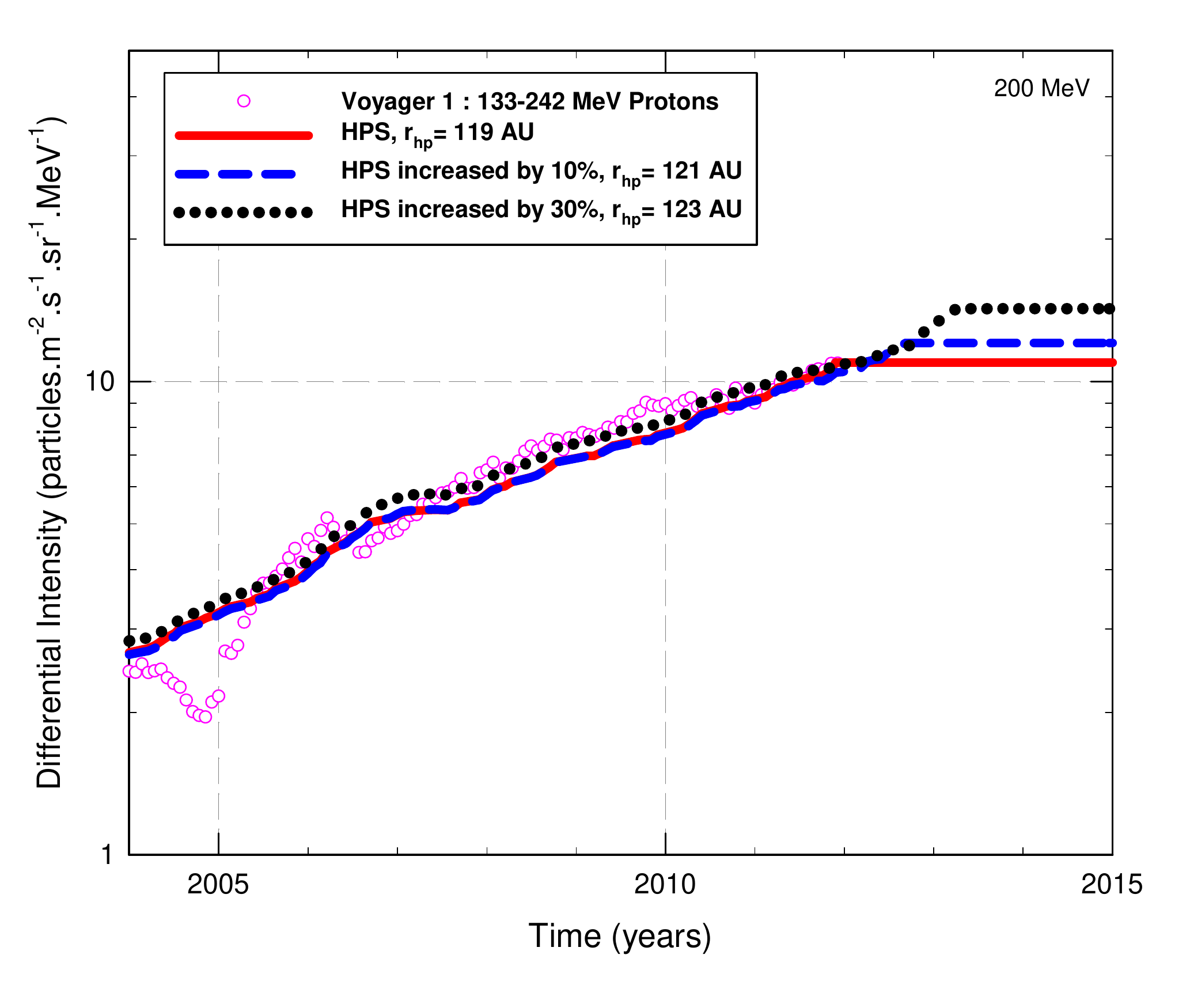}
  \caption{Similar to Figure \ref{figure3} except that here 200 MeV modelling results along the V1 trajectory are shown.}
  \label{figure4}
 \end{figure} 
 
  \begin{figure}[t]
  \centering
  \includegraphics[width=0.48\textwidth]{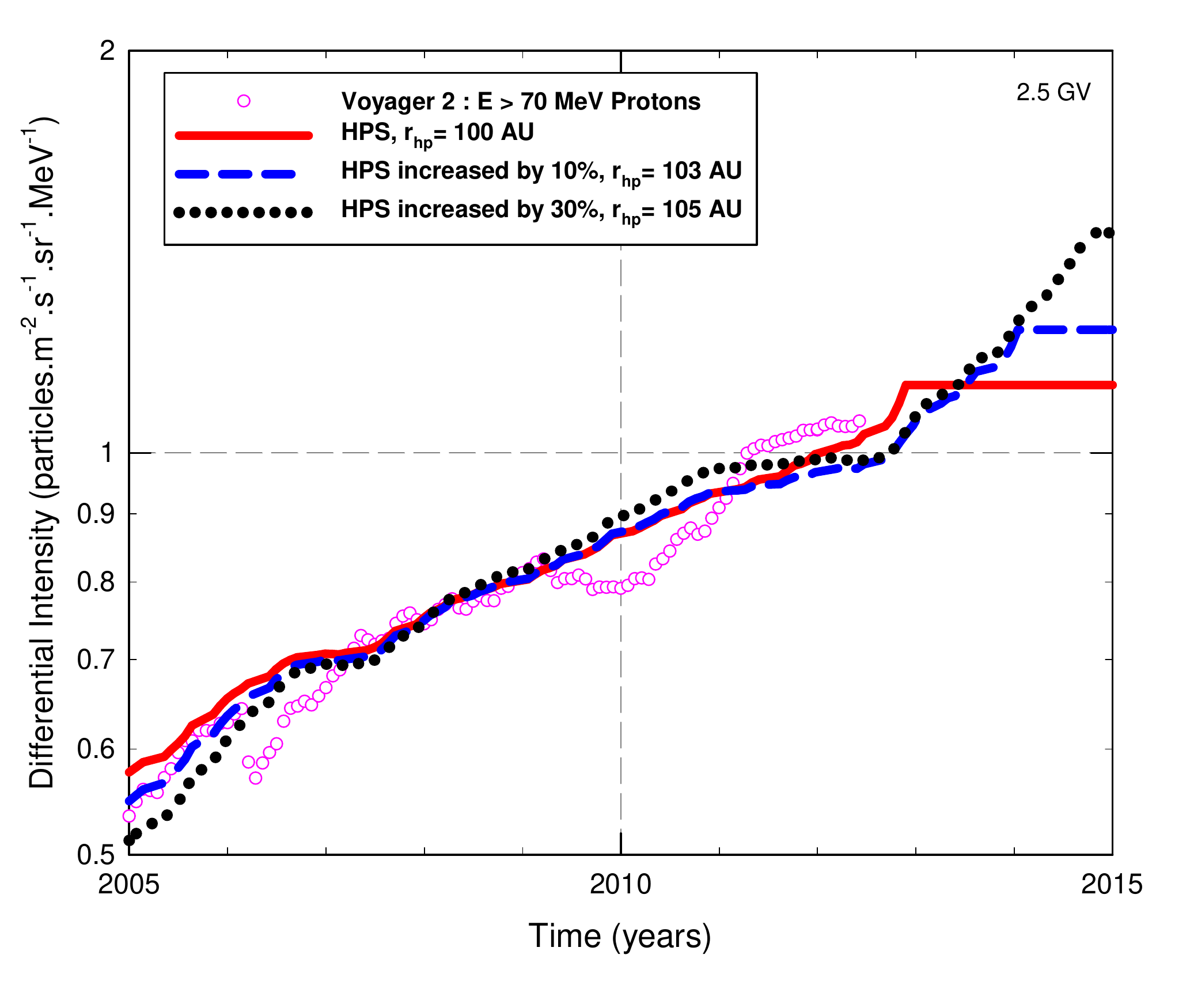}
  \caption{Similar to Figure \ref{figure3} except that here modelling results along the V2 trajectory are shown.}
  \label{figure5}
 \end{figure}
 
A time dependence for the drift coefficient $K_{A}$ is constructed from the theoretical work done by \cite{Minnie07} where $K_{A}$ is scaled with respect to $\delta B$. See \cite{Manuel11} and \cite{Manuel11b} for details.  

The proton spectrum at 119 AU as measured by Voyager 1 is assumed at the HP as the heliopause spectrum (HPS), assuming that no modulation occurs beyond the HP. 

\section*{Results and discussion}
Figure \ref{figure1} shows the model results of 2.5 GV proton at Earth and along V1 \& V2 trajectory compared to IMP 8 (from {http://astro.nmsu.edu}), Ulysses \cite{Heber2009}, V1 and V2 (from {http://voyager.gsfc.nasa.gov}) observations for compatibility. The figure shows that using the recent theories the model successfully simulated long-term CR modulation in the heliosphere at Earth and along both Voyager trajectories on a global scale. 

For the period $\sim$1986--1989 during A$<$0 polarity cycle, protons drift in along the heliospheric current sheet. In this period V2 stayed close to the heliospheric equatorial region and a higher intensity is measured compared to V1 which were at higher latitudes. From the period $\sim$1992--2001 V1 measured higher intensities compared to V2. The model results in Figure \ref{figure1} show that observations at Earth and along V1 can successfully be reproduced by the model on a global scale until 2012. However, along V2 after 2010 the computed  intensities decreased while the observations show an increase possibly due to the assumed symmetrical heliosphere (HP position, $r_{hp}=119$ AU and termination shock position, $r_{ts}=90$ AU) and the same modulation parameters as used along V1 trajectory. This aspect is discussed next.  

Figure \ref{figure2} shows V2 scenarios for a symmetrical and an asymmetrical heliosphere with different diffusion parameters, i.e $a$ values in Equation \ref{perp1}. The computed results for symmetrical heliosphere shows that, even after increasing the $a$ value to 0.03 from 0.022, the model still fails to reproduce the steep increase in CR intensities as observed along the V2 after 2010. This illustrates that any change in diffusion coefficients is not sufficient enough to reproduce the observations when a symmetric heliosphere is assumed. From a thorough parameter study e.g. changing the magnitude, radial and latitudinal dependence of the different diffusion coefficients, the magnitude of the drift coefficient, increasing and decreasing the assumed HPS etc. we came to the conclusion that it is not possible to fit both V1 and V2 observations with the same detail using exactly the same set of parameters in both hemispheres (see also \cite{Ngobeni-2011}). 

However, the scenario in the Figure \ref{figure2} which represents an asymmetrical heliosphere computed compatible CR intensities until 2012 except for the extreme solar maximum periods when model needs some form of merging of the propagated values from Earth \cite{Potgieter92},\cite{Manuel11}. This scenario suggests an asymmetrical heliosphere with different transport parameters in both hemispheres.  Recent theoretical work done by \cite{Opher-etal-2009} and \cite{Pogorelov-etal-2009} suggests a possible asymmetry between the two hemispheres of the heliosphere due to an external pressure resulting from the interstellar magnetic field. Note that asymmetries in internal pressure can also possibly be responsible for such an asymmetry. 

In order to predict  $E$ $>$ 70 MeV and 133-242 MeV proton intensities along the Voyager trajectories we extrapolate the magnetic field magnitude, variance and tilt angle from 2012 onwards up to the time to reach the HP. Scenario 1 (solid line) given in Figure \ref{figure3} is assumed to be the best fit 2.5 GV result along V1 trajectory for assumed HPS at $r_{hp}=119$ AU. The second scenario shows that in order to compute compatible intensities for a 10\% higher HPS the HP must be assumed at 123 AU. The third scenario a 30\% higher HPS at 125 AU computed intensities much higher than the observations suggesting a larger $r_{hp}$, i.e. at 130 AU as shown by black dotted line, which resulted in compatible results when compared to observations on a global scale. This indicate that for an assumed higher HPS the HP position must be increased, or the assumed parameters in the model need adjustment. To decrease the intensities in the inner heliosheath, one can either decrease the magnitude of the diffusion coefficients in this region or increase the assumed modulation boundary, later is done in this study. Similar resuls were obtained for computed 200 MeV protons along V1 as shown in Figure \ref{figure4}. 

Concluding from Figure \ref{figure3} and \ref{figure4} is that within the limitations of this model one cannot learn more about the expected value of the HPS at this energy without knowing the exact location of the HP and value of transport parameters. However, the predicted intensities indicate that V1 should measure on average a steady increase in intensity implying a constant radial gradient based on a significant increase in intensities from current values up to the HP providing that no modulation occurs in the outer heliosheath. This steady increase in intensities may not be necessarily true for V2 which is discussed next. 

Figure \ref{figure5} is similar to Figure \ref{figure3} except that the model results are compared to V2 observations for an asymmetrical heliosphere. The red solid line represents the modelling result with the HPS assumed  as that measured by V1 at 119 AU but specified in the model at 100 AU. This result shows that when an asymmetrical heliosphere with smaller modulation boundary is assumed in the southern hemisphere, the computed intensities produced an improved compatibility with the observations, except for the extreme solar maximum periods when the model needs some form of merging of the propagated values from Earth. The blue dashed line shows the results for a 10\% higher HPS assumed at 103 AU. This scenario also generally reproduced the CR observations along V2 trajectory when a smaller $r_{hp}$ position (103 AU) compared to 123 AU assumed along V1. A third scenario, where a 30\% higher HPS is assumed at 105 AU is shown as black dotted line in Figure \ref{figure5}. This scenario also reproduced the observations on a global scale. All these scenarios predict that V2 spacecraft should measure almost a constant (or decreasing) intensities for some period, where after a sharp increase is expected when it is nearing the HP, similar as along V1 trajectory.

Figure \ref{figure5} shows the difference in intensity profiles predicted by the model up to the HP along V2 compared to V1. Intensity profiles along V1 shows an expected increase in intensities by a constant rate up to the boundary, as shown in Figure \ref{figure3}. However, along the V2 trajectory one can expect a significant difference because of the large distance between this spacecraft and the HP. While we assume a smaller modulation boundary for the V2 calculations shown in Figure \ref{figure5}, these measurements show modulation effects due to the solar cycle. The V2 spacecraft should measure almost a constant (or decreasing) intensity for a few years, where after a sharp increase is predicted when it is nearing the boundary. This study also suggests that without knowing the true HPS or location of HP and transport parameters, one could not make exact predictions of CR intensities along the Voyagers' trajectories. However, possible different scenarios of future CR intensities along these spacecraft could be computed for the assumed different HPS and HP positions.

\section{Summary and conclusions}

Using a time-dependent model, we simulated long-term CR modulation over several solar cycles. Theoretical advances  \cite{Shalchi-etal-2004}, \cite{Teufel-and-Schlickeiser-2002},  \cite{Teufel-and-Schlickeiser-2003} and \cite{Minnie07} in transport parameters are  introduced in the model to computed compatible results at Earth and along both Voyager trajectories. The study revealed that when the same modulation parameters were assumed, which resulted in compatible intensities along V1, the model failed to reproduce the observations along V2. The study shows that any changes in diffusion parameters alone could not reproduce the CR observations along V2 and that changes to the heliospheric geometry were required suggesting an asymmetrical heliosphere. 

The predicted $E$ $>$ 70 MeV and 133-242 MeV proton intensities along V1 indicate that this spacecraft should be relatively close to the HP so that the computed intensities increase with an almost constant rate. However, the predicted model results show that V2 is still under the influence of temporal solar activity changes because of a relatively large distance to the HP when compared to V1. Furthermore, the model predicts that along the V2 trajectory, the intensities may remain generally constant (with temporal effects superimposed) for the next few years and then will start to steadily increase as in the case of V1 observations. 


\vspace*{0.5cm}
\footnotesize{{\bf Acknowledgment:}{This work is partially supported by the South African National Research Foundation (NRF).}}

\end{document}